\documentclass[
preprint,
nofootinbib,
nobibnotes,
floatfix,
]{revtex4-2}
\usepackage{enumitem}
\usepackage{graphicx}
\usepackage{dcolumn}
\usepackage{bm}
\usepackage{placeins} 

\usepackage{amsmath, amssymb, dsfont,amsthm,color}
\usepackage{amsmath, amssymb, dsfont,amsthm,color}

\newcommand{\indi}[1]{\mathds{1}_{#1}}
\newcommand{\moyAng}[1]{ \langle {#1} \rangle } 

\newcommand{\R}{\mathbb{R}}

\newcommand{\nobj}{N_{obj}}

\usepackage[linesnumbered]{algorithm2e}
\RestyleAlgo{ruled}
\SetKwInput{KwData}{Input}

\definecolor{darkgreen}{rgb}{0.1,.8,0.1}
\definecolor{darkred}{rgb}{0.8,.1,0.1}
\definecolor{darkblue}{rgb}{0.1,.1,0.8}

\newcommand{\modif}{}
\newcommand{\modifd}{}

\def\Ecal{\mathcal{E}}
\def\methodname#{Quantization Monte Carlo}

\usepackage{geometry}
\usepackage{graphicx}
\usepackage{float}

\begin{document}

\title{The \methodname{} method for solving radiative transport equations}

\author{Laetitia Laguzet}
 \email{Laetitia.Laguzet@cea.fr}
\affiliation{CEA-DAM-DIF\\ F-91297 Arpajon, France}%

\author{Gabriel Turinici}
\email{Gabriel.Turinici@dauphine.fr}
\homepage{https://turinici.com}
\affiliation{CEREMADE, Université Paris - Dauphine - PSL \\ 75016 Paris, FRANCE
\\ \  \\ {Corresponding author: Gabriel Turinici} \\ Gabriel.Turinici@dauphine.fr \\ \ \\}

\date{\today}

\begin{abstract}
We introduce the \methodname{} method to solve thermal radiative transport equations with possibly several collision regimes, ranging from few collisions to massive number of collisions per time unit. For each particle in a given simulation cell, the proposed method advances the time by replacing many collisions with sampling directly from the escape distribution of the particle. In order to perform the sampling, for each triplet of parameters (opacity, remaining time, initial position in the cell) on a parameter grid, the escape distribution is precomputed offline and only the quantiles are retained. The online computation samples only from this quantized (i.e., discrete) version by choosing
a parameter triplet on the grid (close to actual particle's parameters) and returning at random one quantile from the precomputed set of quantiles for that parameter.
We first check numerically that the escape laws depend smoothly on the parameters and then implement the procedure on a benchmark with good results. 
\end{abstract}


\maketitle

\section{Introduction and motivation} \label{sec:introduction}

The time dependent thermal radiative transport equations couple a transport equation with a internal matter density evolution equation. 
Simulating this dynamics is extremely time consuming and 
we are interested in the stochastic (Monte Carlo) approaches and more precisely in the situation involving a large range of 
opacities~; in such cases the particles used in the Monte Carlo simulation will undergo a wide range of behaviors~: on one hand long-time rectilinear propagation interrupted by rare scattering events and on the other hand high intensity scattering with negligible  overall displacement; but all other intermediate regimes are also present.
The two extreme regimes can either be simulated directly or with good quality approximations and the corresponding works have been documented in the literature. 
But treating all regimes simultaneously
has been a challenge and our contribution introduces a unified method to tackle this circumstance. To this end we exploit a hidden smoothness in these models which is situated at the level of the statistics of the escape laws of a particle from a given domain. 

We present briefly the principles of the Monte Carlo method 
used to solve the transport equation~\eqref{eq:transport} and the problem of the 
diffusion limit in a general setting. We then present the state of the art of the methods 
that treat this high collisions regime.

The physical systems that can be simulated through  
the equations below
range from 
general photon transport in radiative hydrodynamics~\cite{pomraning1983equations} 
to  radiation transportation in the dense plasma associated with the inertial
confinement fusion \cite{review_Vlasov_Fokker_Planck_inertial_confinement12,ATZENI2005153} and to  
atmospheric radiative transfer models~\cite{atmospheric_radiative_transfer_code_arts_v2_2011}
including applications in astrophysics \cite{noebauer_monte_2019}; see also \cite{castor2004radiation,mihalas1999foundations,zeldovich1966physics} for further applications.

Consider the  integro-differential transport equation~:
\begin{equation}
	\frac{1}{c} \partial_t u(t,x,\omega) + \omega \cdot \nabla u(t,x,\omega) + (\sigma_a(t,x) + \sigma_s(t,x)) u(t,x,\omega) = \sigma_s(t,x) \moyAng{u}(t,x)
{\modif +s(t,x),}
 \label{eq:transport}
\end{equation}
with time $t \in \R^+$, position $x\in \mathcal{D} \subset \R^d$ ($d\ge 1$ is the dimension), $\omega \in \mathcal{S}^{d}$ (unit sphere in $\R^d$) the angle of propagation and $\moyAng{u}(t,x) = \frac{\int_{\mathcal{S}^d} u(t,x,\omega') d \omega'}{\int_{\mathcal{S}^d} 1 \cdot d \omega'}$
the angular average of $u$ on $\mathcal{S}^d$ {\modif and source term $s(t,x)$.}
The  model describes standard heat radiative transfer equations 
with isotropic scattering 
that can in principle be use for photons and neutrons too~; in the numerical results in section~\ref{sec:transfert_rad} we use photons.

The absorption opacity  $\sigma_a$ and the scattering opacity $\sigma_s$ are (known) functions depending on the spatial discretization. To solve this equation we focus on the  approaches described in~\cite{lapeyre1998methodes} which interpret~\eqref{eq:transport} as a time-evolving probability density and simulate the underlying stochastic process.

When  $\sigma_s(t,x) \to \infty$ we are in the "diffusion limit" and the cost of the Monte Carlo is prohibitive~\cite{densmore2005discrete} : each particle undergoes a high number of collisions with the mean time between two collisions being 
$O\left(\frac{1}{\sigma_s}\right)$. But the asymptotic analysis~\cite{larsen1980diffusion} shows that  \eqref{eq:transport}  converges towards the diffusion limit equation~:
\begin{equation}
	\frac{1}{c} \partial_t \moyAng{u}(t,x) = \nabla \cdot \left[ \frac{1}{3 \sigma_s} \nabla \moyAng{u}(t,x) \right] - \sigma_a \moyAng{u}(t,x).
	\label{eq:diffusion} 
\end{equation}

The equation $\eqref{eq:transport}$ appears in particular when solving radiative transfer equations where an isotropic scattering term is necessarily added by the  \textit{Implicit Monte Carlo} linearization method \cite{FLECK1971313} in order to artificially represent the phenomena of absorption and re-emission.
Another approach to avoid artificial scattering is proposed in \cite{brooks1989symbolic} but the problem remains unchanged when important physical scattering terms are present. In the context of radiative transfer, several methods have been proposed exploiting the limit regime~\eqref{eq:diffusion}.

The \textit{Random Walk} (RW) methods \cite{fleck1984random, giorla1987random} exploit the fact that the trajectories of the particles are close to those of a Brownian motion:  in an optically thick medium they replace (a part of ) the trajectory by a single diffusion step in the largest sphere contained in the mesh.
The \textit{Random Walk} methods have the advantage to activate everywhere on the domain {\modifd because it is not depending on the position of the particle but on others attributes of a particle (like its frequencies that may determine its opacity for multi-group simulation)}, and are easily applied to $3$-dimensional problems as well as multi-group problems. Their use in a production context remains limited by their strong dependence on mesh size (the smaller the mesh size, the smaller the sphere where the method will be applied) and the loss of precision introduced by the use of the diffusion limit for transient regimes.

Initially called \textit{Implicit Monte Carlo Diffusion}, the \textit{Discrete Diffusion Monte Carlo} (DDMC) method~\cite{gentile2001implicit,densmore2005discrete,densmore2006hybrid} splits the domain into two regions: one optically thick region solved by a Monte Carlo method using a diffusion equation and another part treated by the \textit{IMC} method. The numerical simulation in the optically thick region uses a linearization similar to the \textit{IMC} method. A new type of particle is then introduced to solve the diffusion equation. The advantage of this method is that it does not have any net flux to consider between the diffusion and transport regions (the flux is carried by the particles) and the particles can go from one region to another (by a conversion) and, more importantly, can change the cell (having different $\sigma_a$ and $\sigma_s$ values) with no particular treatment. The introduction of a new type of particles to treat the diffusion region allows easy
treatment of the interface between the transport and diffusion regions. Contrary to the  \textit{RW} methods, the efficiency of these methods is not dependent on the mesh; however their use is still restricted by the loss in precision introduced by the diffusion approximation when particles change the region.

Hybrid approaches \cite{pomraning1979, clouet2003hybrid} solve the diffusion equation analytically in some spatial areas and use the \textit{IMC} approach in others. Both methods are coupled by boundary conditions. The hybrid methods use an analytical resolution of the scattering equation when certain criteria are met (delimited areas or according to the frequency group). The use of these methods remains limited by the coupling between the analytical resolution of the diffusion equation and the Monte Carlo method solving the transport equation which is delicate as well as the choice of criteria (e.g. the definition of areas where the diffusion approximation can be used).

{\modif 
The hybrid approaches have been included  in the multi-scale paradigm proposed by Coelho et al. \cite{COELHO201636} that 
also contain domain decomposition strategies and micro–macro models; this 
allows to deal with transport, intermediate and diffusive regimes and are applied to 3D transient problems with collimated radiation.}

When the coefficient $\sigma_s(t,x)$ in \eqref{eq:transport} is large, the classical Monte Carlo method uses Markov particles that undergo an important number of scattering events. The randomness of the scattering part dominates and after a certain time the state of the particle follows  a probability law; in this case the  \textit{RW} approximation is justified. However, there are always intermediary regimes when the number of collisions is big enough to slow down the computation but 
  does not meet the necessary threshold to warrant the use of the diffusion approximation.

A new Monte Carlo method that is efficient regardless of the value of $\sigma_s$ and that does not reduce the accuracy of the solution is still a challenge. Ideally the method should  not be sensitive to the mesh used (i.e. robust to the change in value of $\sigma_s$ and not limited to simple spatial domain e.g., a sphere); 
and it needs to be valid regardless of the value of $\sigma_s$ (or that activates according to criteria independent on a choice of spatial areas such as methods of \textit{RW} type)  irrespective of whether the diffusion approximation is valid or not.

Our approach, called the \methodname{} method, is to not use the diffusion limit approximation but to work with an approximation of the probability law of the \textbf{exact solution} of the escape time, position and direction from the spacial cell.

 This involves an offline-online approach to construct a procedure to sample from the escape probability distribution. Comparable strategies have been used previously in radiative transfer equations under the name of 
 look-up tables \cite{stamnes1981new,mobley2005interpretation,lyapustin2011multiangle,martino2017automatic}; look-up tables are used to precompute and store numerical solutions for various sets of input parameters~; the look-up table is queried during the online simulation to retrieve (using interpolation if necessary) the appropriate pre-computed solution for the current input parameters. 
 Note however that our output is not a single (possibly vector-valued) object but a probability law conditioned on some input parameters. While our technical approach, is different the similarity with look-up tables is to use the precomputation effort to reduce the online cost.

The outline of the paper is the following~: we describe in section \ref{sec:method} our method based on an offline-online approach that exploits the quantiles of the escape laws from a domain. 
The assumptions of the method are checked numerically in section \ref{sec:test_exit_time}
and then the method is tested on a benchmark with good results in sections~\ref{sec:transfert_rad}.  
Concluding remarks are presented in section~\ref{sec:conclusion}.

\section{The \methodname{}  method} \label{sec:method}
 We will consider $d=1$ in all  this section and work on a segment (eventually divided in several sub-intervals). To ease notations we will also use $\sigma$ instead of the scattering opacity $\sigma_s$.

\subsection{Toy model illustration}

We recall here a simple example used later in the numerical tests in section~\ref{sec:transfert_rad} and that will be useful to describe the \methodname{} method below.
This approximation corresponds to a $S_2$
discrete-ordinates method
 (see \cite[section 16.3 page 502]{modest2021radiative} reated to the Schuster-Schwarzschild equations \cite[section 14.3 p 456]{modest2021radiative} and is a standard benchmark for applications we envision.
Consider a $1D$ particle  in the segment $[x_{min},x_{max}]$ situated at the initial time $t=0$ at position $x=x_{init}$ with angle $a\in \{-1,1\}$. The total remaining simulation time is $t_{max}$; in the general simulation $t_{max}$ equals the overall time step $\Delta t$ decremented by any previous time increments for this particle (for instance when the particle traverses several cells during the same $\Delta t$).

 The exact evolution of the particle is the following:  rectilinear movement in direction $a$  for a time $\tau$ (exponential random variable of mean $1/\sigma$) then a collision takes place. This collision changes the angle uniformly at random to a new value  $a'\in\{-1,1\}$. Then the process repeats until either   boundary is reached~: $x=x_{min}$ or $x=x_{max}$ or $t=t_{max}$.

\begin{figure}[htb!]
	\begin{center}
		\includegraphics[width=.49\textwidth]{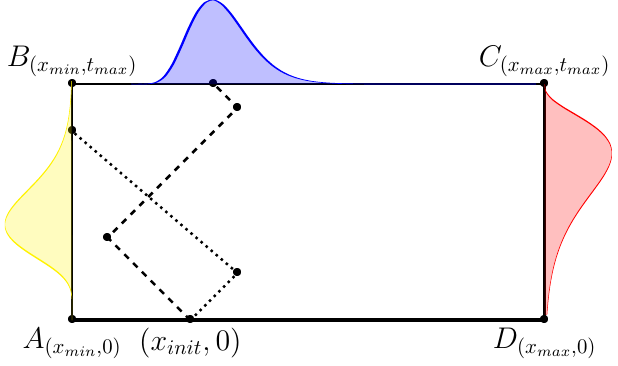}	
		\caption{\small \it An illustration of the escape dynamics of a particle starting at $x_{init}$ and undergoing collisions after $Exp(\sigma)$ time (exponential random variable of average $1/\sigma$). The particle can escape through any of the domain's frontiers: either because it escapes the spatial domain (dotted trajectory) or because the time is up (dashed trajectory). The random events accumulate into a probability law 
	denoted	$\Ecal(\sigma,\ell,t_{max})$
			with support on the boundaries of the time-space domain (together with a escape angle direction attribute).
		}  \label{fig:xminxmaxtmax}
	\end{center}
\end{figure}

We are interested precisely in this escape place (one of the extremities of the segment or of the time domain) and the escape angle. This is a random variable whose distribution will be denoted 
$\Ecal(\sigma,\ell,t_{max})$
where $\ell=(x_{init}-x_{min})/ (x_{max}-x_{min})$ is the relative initial position of the particle.
An illustration is given in figure~\ref{fig:toy_example} for general values of $x_{min}$ and $x_{max}$. We explain below for the case $x_{min}=0$, $x_{max}=1$, the general situation being just a rescaling.	
Note that the possible values of the escape (random) variable are triplets consisting of an escape position, an escape time and an escape angle ; the angle is restricted by the $S_2$ model to be either $-1$ or $+1$. The support will therefore be a subset of $\R^3$ but it is very sparse subset and can be described as being the union of segments $AB, BC, CD$ in figure \ref{fig:xminxmaxtmax} and, as third value, an angle equal to either $-1$ or $+1$.
Using the the symbol $\bigcup$ for the union of ensembles and $\times$ for the product of ensembles, the mathematical transcription of the support is $ \Big(AB \bigcup BC \bigcup CD \Big)\times \{-1,1\}$. This ensemble can be further restricted because a particle that escaped through the side $AB$ will certainly have an escape angle of $-1$ and a particle escaping through $CD$ side an escape angle of $+1$. Exiting through the $BC$ side imposes no a priori conditions on the particle angle but we can assume that the values of the angle $-1$ and $+1$ are likely  independent of the position, because in the $S_2$ model, the collision will generate uniform sampling of the angle. So the support 
(i.e., the set of all possible values taken) of the escape random variable and of the associated probability law $\Ecal(\sigma,\ell,t_{max})$ is with the notations in figure 
\ref{fig:xminxmaxtmax} :
 \begin{equation}
  \Big( AB \times \{-1\} \Big)  \bigcup \Big(BC \times \{-1,1\} \Big) 
\bigcup \Big( CD \times \{1\} \Big).
 \end{equation}

Note that, although the distribution seems to be $3$ dimensional, conditional on knowing the escape side, only one dimension is essential, for instance escaping through the segment $AB$
leaves only the escape time distribution unknown because the escape position is certainly $x_{min}$ and the escape angle $-1$; escaping through the segment $CD$ is similar ; escaping through $BC$ leaves unknown only the distribution of the escape position because the time is $t_{max}$ and the distribution for the angle is uniform between $-1$ and $+1$ and independent of the position 
(both values $-1$ or $1$ are as likely if at least one collision took place).
%
%
The colored areas in figure~\ref{fig:xminxmaxtmax} are an "artist view" (that is, not corresponding to any specific parameters) of the three "important" conditional distributions: 

- the left area (yellow in color figure) is the distribution of the escape {\bf time} $t$ 
at which the particle reached some $(x_{min},t) \in AB$, 
conditioned by the fact that the particle escaped through the segment $AB$ i.e., reached $x=x_{min}$ before $t_{max}$ and before reaching $x=x_{max}$ ;

- likewise, the top area (blue in color figure) is the distribution of the escape {\bf position} $x$ at which the particle reaches  $(x,t_{max}) \in BC$
conditioned by escaping through $BC$;

- the right area (red in color figure) is the analog of the left area but for the situation of an escape through $CD$.

\subsection{The method} \label{sec:method_algo}

The section \ref{sec:introduction}  highlighted the difficulty of dealing with the diffusion limit of the equation \eqref{eq:transport} and the limitations of existing Monte Carlo methods. We propose a new Monte Carlo method, inspired by algorithms such as \textit{Random Walk}, that 
works with the probability laws $\Ecal(\sigma,\ell,t_{max})$ of escape from a cell and is based on vector quantization techniques~\cite{book_quantization_measures}.

\begin{enumerate}
\item We define grids of representative values of the main parameters concerned~; for instance in 1D, we employ a grid $G_{sc}$ for $\sigma$ values (in practice a log-uniform grid from $7.5\times 10^{-3} cm^{-1}$ to $9.0\times 10^{6} cm^{-1}$), a grid $G_{time}$ for simulation time values $t_{max}$ (uniform grid from $400fs$ to $40000fs$) and a grid $G_{ini}$ for relative initial position in the cell from $0\%$ to $100\%$ relative to left segment end. Each grid $G_{sc}$, $G_{time}$, $G_{ini}$ has $100$ points. We denote $|G|$ the size of a grid $G$.

\item An \textbf{offline} computation is done once and for all (independent of the final simulation) in order to obtain an approximation of the joint distribution (escape time, escape point, escape direction) 
 $\Ecal(\sigma,\ell,t_{max})$ 
as a probability distribution. For each point in 
$G_{sc} \times G_{time} \times G_{ini}$
 we compute and store the quantiles of the law. This approximation is valid beyond the framework of the diffusion limit, in particular it does not use any analytical form. In practice we perform $1500$ simulations for each point in $G_{sc} \times G_{time} \times G_{ini}$ but extract only a predefined number of quantiles from the whole distribution (cf. previous remarks on the fact that distribution is essentially one dimensional). 
When $\sigma$ is large enough to ensure that the diffusion approximation is valid, one can sample this law using this diffusion approximation. In practice we use a very conservative approach by replacing, for $\sigma$ large, several collisions with one collision provided that the diffusion approximation ensures that the probability to escape is less than $10^{-6}$.
 	The method is detailed in \cite{laguzet_turinici_exit_law24}. 	
 	 Note that this is only a way to compute faster the exact law but the \methodname{} does not depend on this choice, any sampler of the exact escape law will do.
  We will denote by $J$ the number of quantiles used, $J$ is a parameter of the method. In practice we set $J=100$.
The quantiles are minimizers of the Huber-energy distance to the target and correspond to the optimal quantization (i.e., discretization) of the measure~;  when quantizing with $J$ points the optimal quantiles have been proven, cf. \cite[prop. 21]{measure_quantization_turinici22} and \cite[prop. 3 and 4]{turinici_deep_2023} to be the 
 $\frac{j+1/2}{J}$, $j=0,...,J-1$ quantiles.  
This part of the simulation is highly parallelizable. The results are stored as a 
$|G_{sc}| \times |G_{time}| \times |G_{ini}| \times J$ array of escape points $x$ or $t$  
together with the $3$ positive numbers  (summing up to $1$) indicating the probability of escape through each side; for us $J=100$, the number of points is $100^3\times 103$ requiring $\sim 800Mb$ of storage.
{\modif To store the quantiles we proceed as follows: for each point of the grid we sample exit points and obtain an empirical sampling of $1500$ points. Any such point is either situated on the segment $AB$ in 
figure~\ref{fig:xminxmaxtmax}
(in which case the particle exited through the left side), or on the segment $BC$ in figure~\ref{fig:xminxmaxtmax} (the particle remained in the interior until the time was up) or on the segment $CD$ (the particle exited through the right). We map now the curve AB-BC-CD into the segment $[0,1]$ by a piecewise linear mapping: $AB$ is mapped into $[0,p_L]$ where $p_L$ is the percentage points that exited through the left side. Segment $CD$ is mapped to $[1-p_R,1]$ where $p_R$ is the percentage of points that exited through right side and $BC$ is mapped to $[p_L,1-p_R]$. We obtain a empirical distribution on $[0,1]$ having $1500$ points, from which we take 
the quantiles of $\frac{j+0.5}{100}$ for $j=0,...,99$. These are mapped back on the curve $AB-BC-CD$ and stored in the dataset.
}
\item During the {\bf online} simulation, 
each time that a particle 
of parameters $(\sigma,\ell,t_{max})$ 
needs to be advanced to its next escape point, 
a set of parameter values 
$\sigma^g,\ell^g,t_{max}^g$ from the 3D-grid 
$G_{sc} \times G_{time} \times G_{ini}$ is chosen (see below for details) and 
 a random quantile from the stored distribution 
 $\Ecal(\sigma^g,\ell^g,t_{max}^g)$ is selected 
 and returned to the user. The particle is advanced with the corresponding space/time increments prescribed by the escape quantile returned. 
The grid point 
$\sigma^{g},\ell^g,t_{max}^g$ is chosen by identifying, for each of the parameters 
$\sigma,\ell,t_{max}$ 
 the $2$ closest values of the grid~: 
$\sigma \in [\sigma^{k_1},\sigma^{k_1+1}]$,
$t_{max} \in [t_{max}^{k_2},t_{max}^{k_2+1}]$,
$\ell \in [\ell^{k_3},\ell^{k_3+1}]$~; then we select one of them at random with probabilities depending on the relative distance between the actual parameters and the grid points, for instance $\sigma^g=\sigma^{k_1}$ with probability 
 $(\sigma^{k_1+1}-\sigma)/ (\sigma^{k_1+1}-\sigma^{k_1})$. {\modif The function thus obtained is called {\it computeWithQuantization} and will be used in algorithm~\ref{algo:QIMC}.}
 
\end{enumerate}	
Such an approach does not raise questions of validity of the diffusion limit or of the calculation of the escape time from the spheres (which resort to partial differential equations with assumptions and boundary conditions sometimes difficult to tackle  cf.~\cite{https://doi.org/10.1002/eng2.12109, giorlaCEAn84}).

The method is called "quantized" because we always sample from a discrete, pre-defined list of quantiles. In practice this dimension of quantization is not any more surprising than, e.g. space discretization of the mesh and if enough quantiles are considered the contribution to the overall error is negligible. The foundations of the method are well established 
(see~\cite{book_quantization_measures} for general information on the mathematical objects and \cite{measure_quantization_turinici22} more specifically tailored to our applications).
{\modifd Note that a specific drawback of the quantization is that, since particle is moved directly from initial position in the cell to exit position, 
we do not have access to events that occurred before exit and all events are agglomerated at exit time. 
}
\section{Numerical tests}

\subsection{Toy model tests: escape time and position} \label{sec:test_exit_time}

\begin{figure*}[htb!]

\hbox{\hspace{5.9cm} Exited left \hspace{0.9cm} \parbox{1.6cm}{\linespread{1}\selectfont No exit before final time} \hspace{0.9cm} Exited right}

\vspace{0.5cm}

\raisebox{1.5cm}{$\sigma=0.75$ \hspace{0.5cm}}
\includegraphics[width=0.49\textwidth]{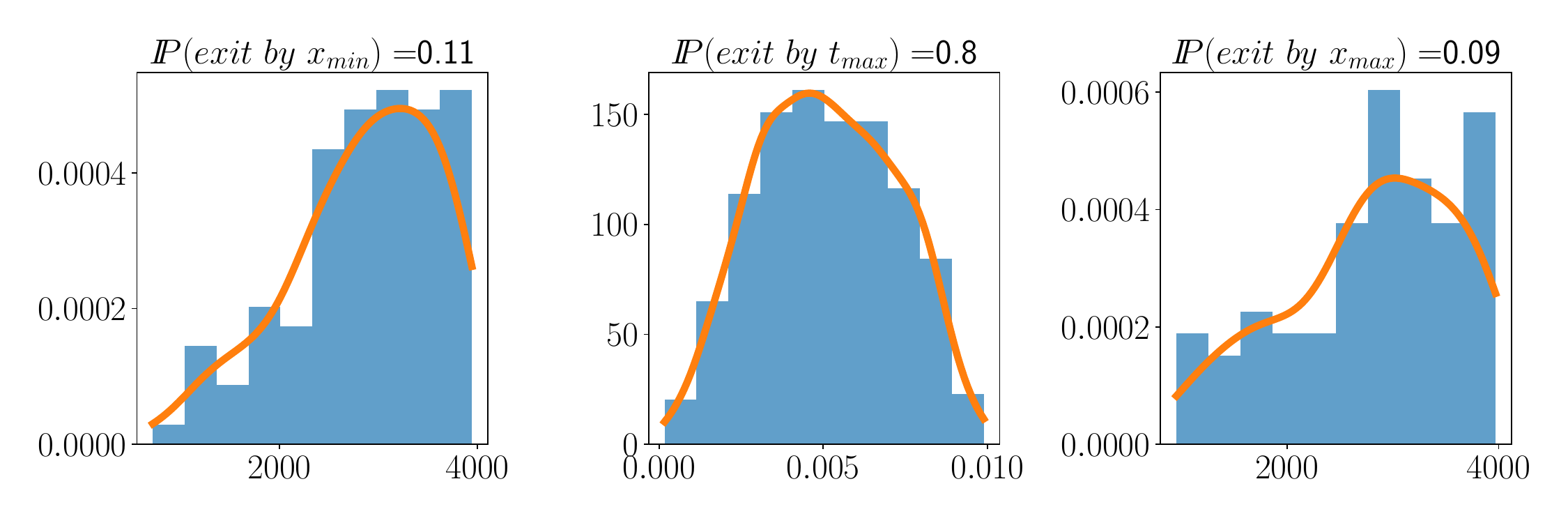}

\raisebox{1.5cm}{$\sigma=1.0$ \hspace{0.5cm}}
\includegraphics[width=0.49\textwidth]{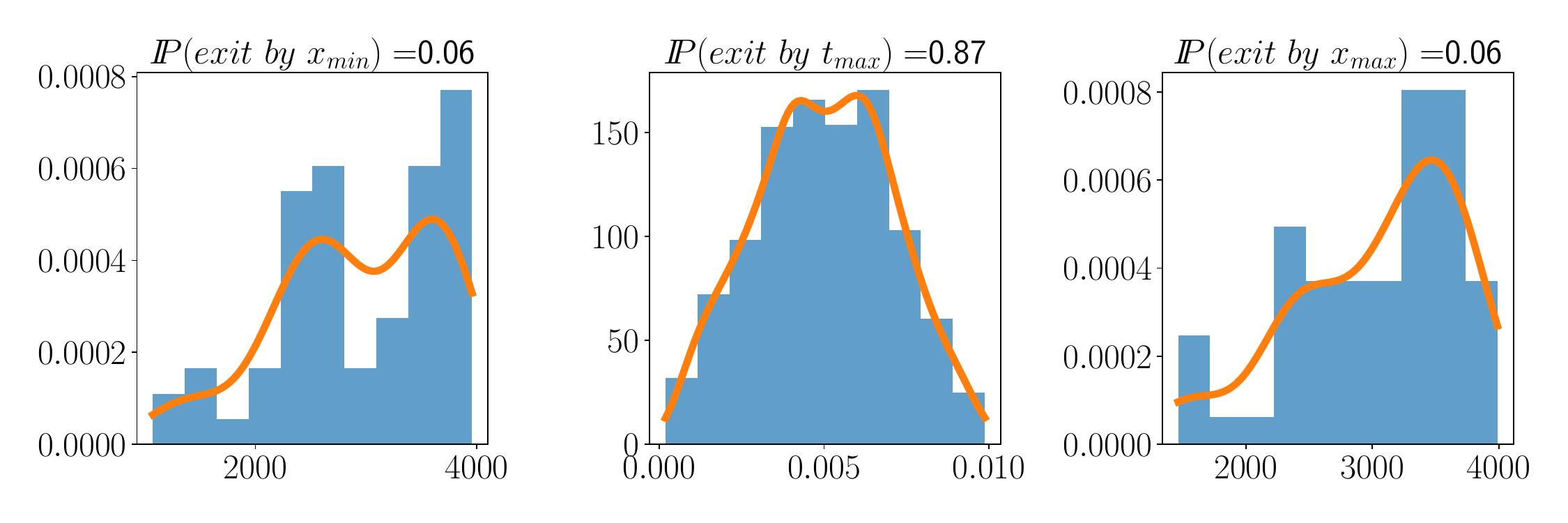}

\raisebox{1.5cm}{$\sigma=1.25$ \hspace{0.5cm}}
\includegraphics[width=0.49\textwidth]{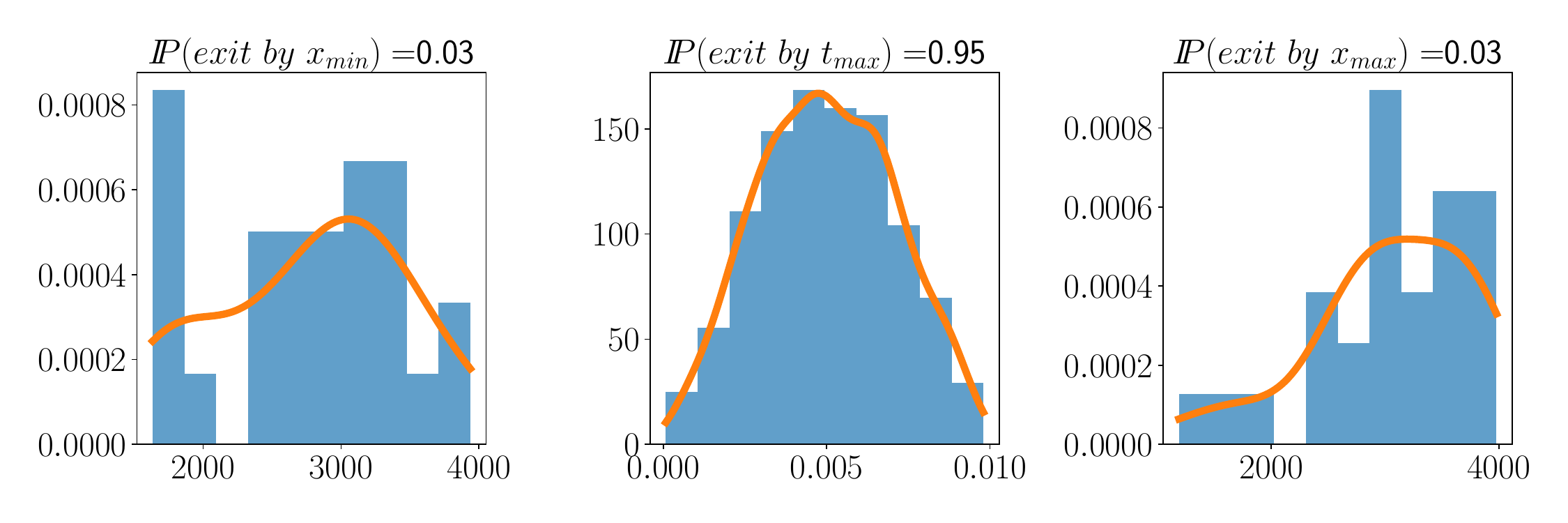}

\raisebox{1.5cm}{$\sigma=7.5$ \hspace{0.5cm}}
\includegraphics[width=0.49\textwidth]{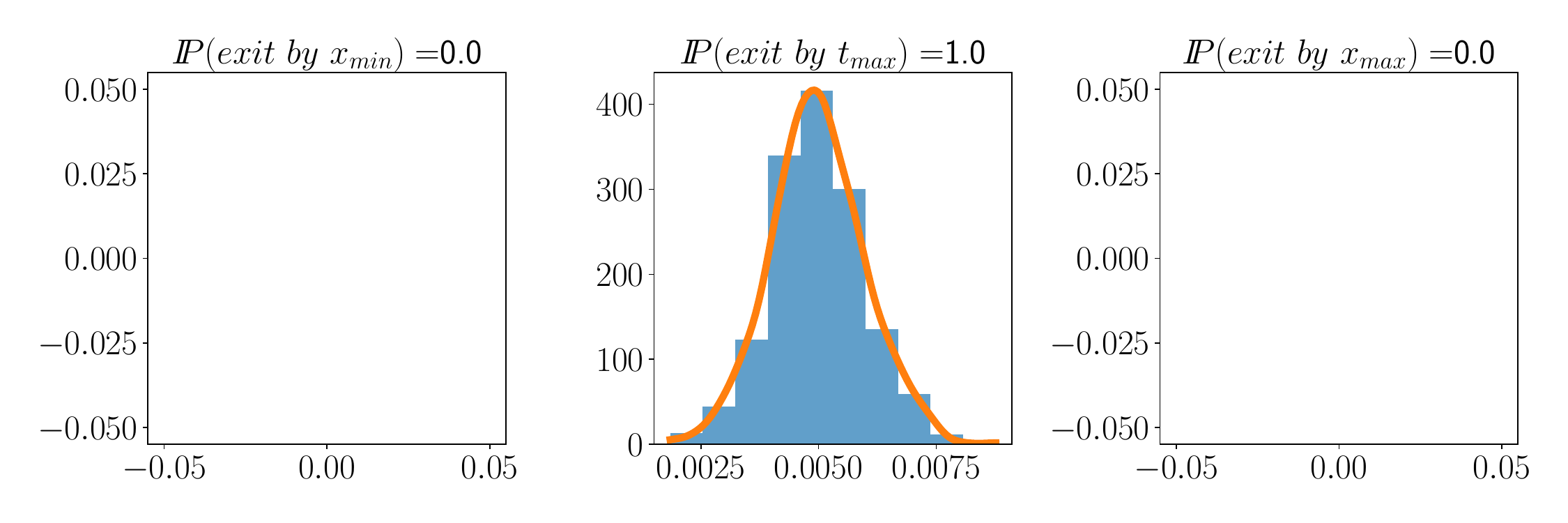}

\raisebox{1.5cm}{$\sigma=10.0$ \hspace{0.5cm}}
\includegraphics[width=0.49\textwidth]{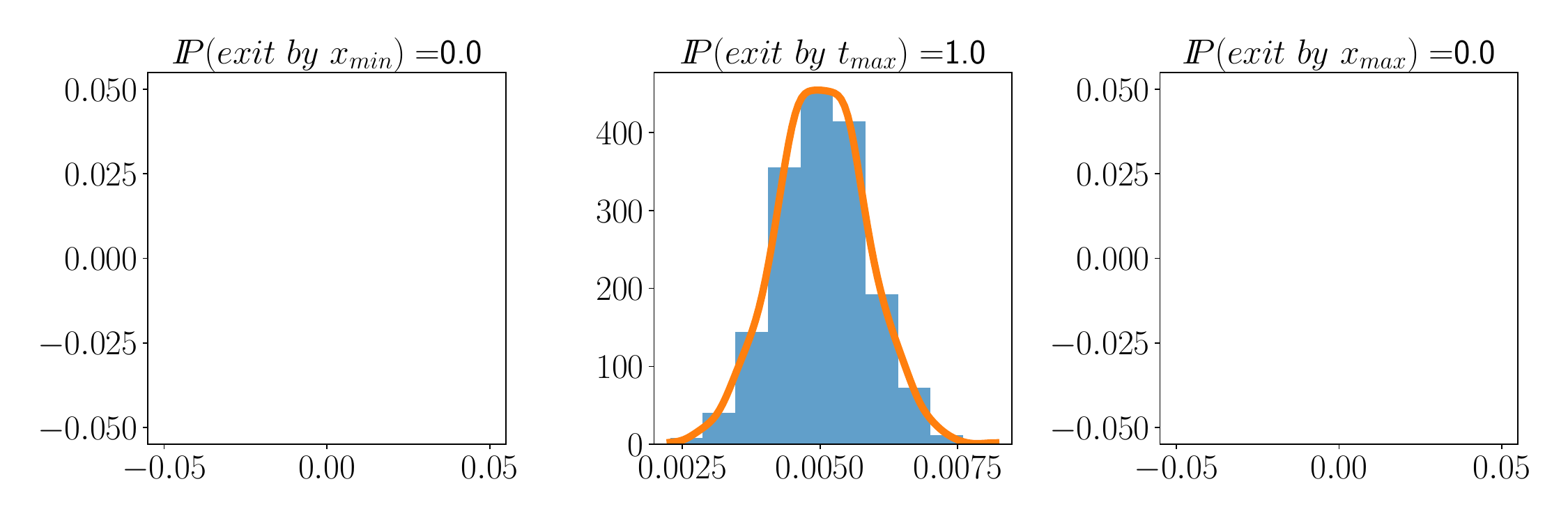}

\raisebox{1.5cm}{$\sigma=12.5$ \hspace{0.5cm}}
\includegraphics[width=0.49\textwidth]{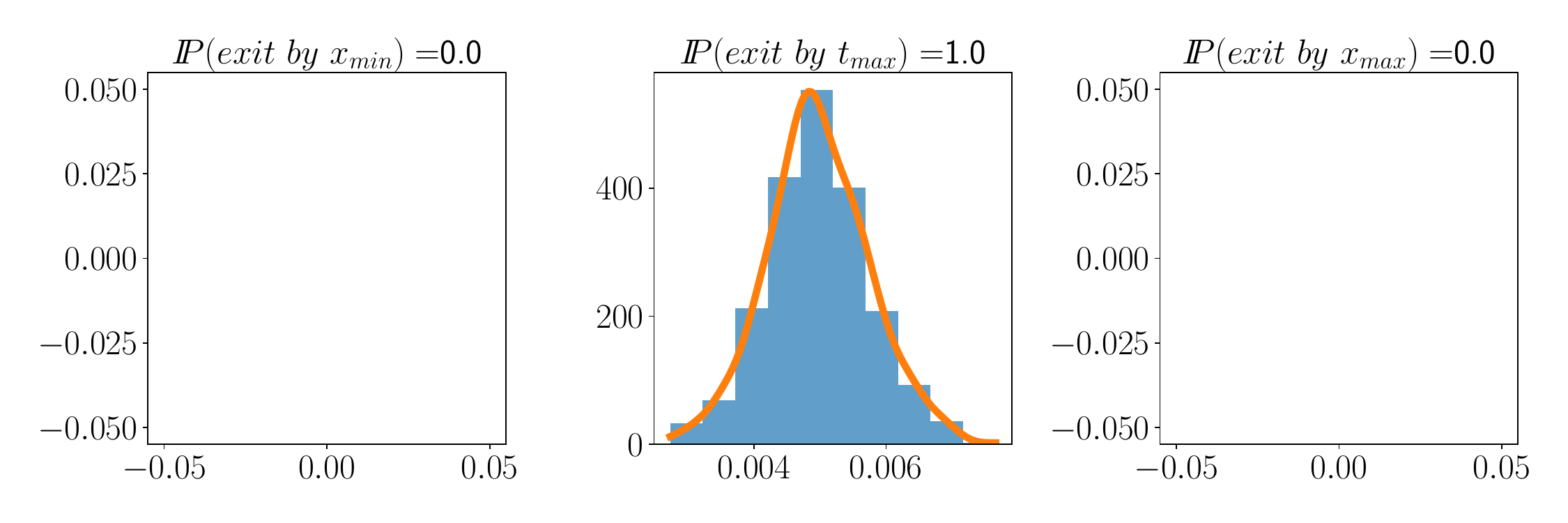}
\caption{\modif \small  \it Escape histograms from the time-space domain 
	$[x_{min},x_{max}]\times [0, t_{max}]$ with $x_{min}=0.0$, 
	$x_{max}=0.01$, initial direction $+1$, 
	$t_{max}=4000 fs$, speed $3.0\times 10^{-5}$ (speed of light in fs/cm), $x_{init}=0.005$.
The orange lines are kernel density estimation of the blue distributions; these are put for convenience and not used in any way in the computations.
	The probabilities of escape are given in the title of each plot.}
\label{fig:toy_example}
\end{figure*}

In order for the \methodname{} method to work conveniently, one needs to ensure that 
the distribution  $\Ecal(\sigma,\ell,t_{max})$  is close to the mixing of the closest  distributions $\Ecal(\sigma^g,\ell^g,t_{max}^g)$ on the grids. This, at its turn, depends on the 
smoothness of the mapping $(\sigma,\ell,t_{max}) \mapsto \Ecal(\sigma,\ell,t_{max})$ that we investigate in the following. More precisely, we plot in figure~\ref{fig:toy_example} several histograms corresponding to different typical parameter values encountered in the numerical tests in section~\ref{sec:transfert_rad}. 
{\modif 
We take	$x_{min}=0.0$, 
	$x_{max}=0.01$, initial direction $+1$, 
	$t_{max}=4000 fs$, speed $3.0\times 10^{-5}$ (speed of light in fs/cm), $x_{init}=0.005$ and change the $\sigma$ parameter (in $cm^{-1}$) : 
 $\sigma=0.75$ (first row of plots),   
 $\sigma=1$ (second row), $\sigma=1.25$ (third row)
  $\sigma=7.5$ (fourth row)
  $\sigma=10$ (fifth row) and  $\sigma=12.5$ (sixth row of plots).
   Note that in the first and third column the relevant information, that is the abscissa, is a time between $0$ and $t_{max}$ and in the middle column the relevant information is a space position between $x_{min}$ and $x_{max}$.
  For instance the histogram at line $2$ column $3$ corresponds to a test with $\sigma=1$ and presents the histograms of the exit time for particles that have exited through the right side (this is a conditional law).
  The probability for a particle to exit through the right side is $0.06$ and is given in the title of the plot. In this case most particles remain in the domain till the final time because the probability to exit before the time is up is $0.87$ (title of the histogram in line 2 column 2). This is even more so when $\sigma$ is large (last three rows) : the collisions are too many and the particle does no significantly move i.e., it only escapes because the time is consumed. For instance in row $4$ column $1$ (for $\sigma=10$), it should be plotted the histogram of the distribution of the time  values when particles exited through left side. But in this particular case no particle exit through left side (we read this in the title where probability is indicated to be $0.0$) so there is no histogram to plot.}
As expected, the laws vary slowly with the parameters.
For instance, in practice we noted that a grid of values for $\sigma$ spaced log-uniform by about $25\%$ increase from one point to another gives very satisfactory results.

\subsection{Propagation of a Marshak-type wave with multi-regime physics and temperature dependent opacity}\label{sec:transfert_rad}

\begin{figure*}[htb!]
	\includegraphics[width=0.9\textwidth]{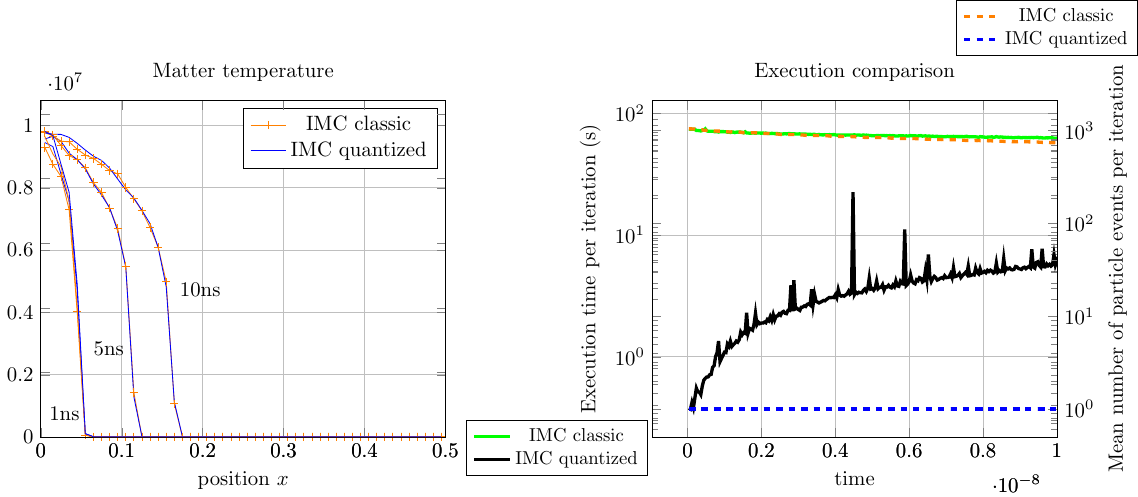}
	\caption{\small \it Results of the simulation described in~\ref{sec:transfert_rad} (multi-regime physics, temperature dependent opacity). The number of events per particle for the \methodname{} is reduced with respect to  the  reference while  keeping the physical properties of the solution. 
 {\bf Left image~:} temperature profile for the times $1ns$, $5ns$ and $10ns$.
{\bf Right image dashed lines, right axis ~:} mean number of particle events per iteration  for the classical Monte Carlo trajectory compared with the quantized simulation; {\bf Right image solid lines, left axis :} execution time per iteration for the two procedures. All plots refer to the same simulations.}
	\label{fig:second_result_M}
\end{figure*}

We test the method on the propagation of a Marshak-type wave in an opaque medium (see \cite{marshak1958,MCCLARREN20089711} for details) which is considered a good benchmark for difficult multi-regime computations. {\modif This test case corresponds to an initially cold material with radiation incident on the surface (here at the left of the spacial domain). At the beginning, the material is cold, with a high opacity (see the formula of the opacity below in table \ref{tab:valeursMarshak}). When the wave is entering in the material (at about $10 ns$) high and small opacities are presents in the simulation due to the dependence of the opacity to the material temperature. 
}
We assume an ideal gas equation under the gray approximation.
The Monte Carlo method used here is based on the Fleck and Cummings  linearization (see \cite{FLECK1971313} and \cite{laguzet_turinici_exit_law24} {\modif for all details but we recall the main idea below}).
{\modif 
The purpose of this linearization is to solve the joint equations describing the particle transport {\bf and} the evolution of the matter temperature. 
The evolution of the matter temperature is coupled with the transport equation \eqref{eq:transport} that provides the radiative intensity. We obtain a system of two non linear coupled equations: first equation is \eqref{eq:transport} and the second one is~:
\begin{equation}
C_V \partial_t T_{matter} = \sigma_a(T_{matter}) ( \moyAng{u}(t,x) - ac T^4_{matter} ).
\label{eq:matiere}
\end{equation}
To solve this system, we use the Fleck and Commings linearization that assume a space and time discretization. The main idea of the linearization is to introduce a coefficient denoted $\beta$~: 
$$\beta = 
\frac{C_V}{\frac{\partial(a T_{matter}^4)}{\partial T_{matter}}}
 = \frac{C_V}{4 a T^3_{matter}}.$$ 
 Then the coupled matter equation can be rewriten as: 
\begin{equation}
    \frac{\partial (a T_{matter}^4)}{\partial t} = \sigma_a \beta ( \moyAng{u}(t,x) - ac T^4_{matter} ).
    \label{eq:matiere2}
\end{equation}

As we suppose the local thermodynamic equilibrium the emission source in \eqref{eq:transport} is  $\frac{acT_{matter}^4}{4 \pi}$. To estimate the matter temperature for the iteration $[t_n, t_{n+1}]$, we fix the opacity (noted $\sigma_a^n$) using $T_{matter}(t_n)$ and we integrate in time \eqref{eq:matiere2} to obtain the following approximation, for time $t_n$ to $t_{n+1}$: 
\begin{equation}
    aT^4_{matter}(t) = f^m a T_{matter}^4(t_n) + \beta_n \sigma_a(t_n) (1-f^m) (t_{n+1}-t_n)\moyAng{u}(t)
    \label{eq:aT4}
\end{equation} with 
\begin{equation}
    f^n = \frac{1}{ 1+ \sigma_a^n (t_{n+1}-t_n) \beta^n c} \textnormal{ and } \beta^n = \frac{4 a T_{matter}^3(t_n)}{C_V}.
    \label{eq:fleck_beta}
\end{equation}
We introduce \eqref{eq:aT4} in the equation \eqref{eq:matiere} to obtain:
\begin{equation}
C_V \partial_t T_{matter}(t) = \sigma_a(T_{matter}(t_n)) f^n ( \moyAng{u}(t,x) - ac T^4_{matter}(t_n) ).
\label{eq:matiere_linearise}
\end{equation}
The Monte-Carlo method allows to obtain $\moyAng{u}$~; using the equation \eqref{eq:matiere_linearise} we obtain the matter temperature at $t_{n+1}$.

The linearization will induce a source term in the equation. Together with the boundary condition (the left side has incident radiation) this requires, for a proper treatment, to create new particles at each time step. This is the "emission" part of the algorithm.
}
%
The emission method is the same for the two classical Monte Carlo method and our method~: 
	for the photons belonging to the input flux the 
	starting point is at 0cm while for the others the starting point is chosen uniformly on the cell.  
	In the particle displacement step, the absorption is treated by the exponential decrease of the weight of a particle as it is classical, see description in \cite[chap. 3]{lapeyre1998methodes}; note that this is not a scattering type event. Moreover, if the weight of the particle is below some predefined  value (computed with the cell based sampling method {\modif \cite{laguzet2020}}), the particle is absorbed in the current cell and the numerical treatment stops for this particle. {\modif We describe in the algorithm \ref{algo:IMC} the classical IMC method used in this paper to compute the reference solution in figure~\ref{fig:second_result_M} and in algorithm \ref{algo:QIMC} the quantized version we introduce in this work. The algorithm \ref{algo:all_mc} gives all the details for this test case.}

\begin{algorithm}
{\modif 
\KwData{
$x_p(t)$ is the position of the particle, $\omega_p(t)$ the angle of the particle, $t_p$ the remaining life time for the particle and the $\alpha_p(t)$ its weight. The particle is on cell $m$; this cell is represented by a segment $[x_{min}, x_{max}]$ ; the attributes $\sigma_a^m$ and $f^m$ of the cell are known from the table of mesh sizes $MCestimate$.}
Assign $\omega = \omega_p(t)$, $x = x_p(t)$, $\alpha = \alpha_p(t)$\;
\While{$t_p > 0$}{
Compute $d_{iter} = c \times t_p  $, $d_{exit} = (x_p(t) - x_{min}) \indi{\{  \omega = -1 \}} + (x_{max} - x_p(t)) \indi{\{\omega =+1\}} $ \;
Draw $u_{01} \sim \mathcal{U}_{\{[0,1] \}} $ and compute $d_{si} = -\log{(u_{01})}/({\sigma^m_a(1-f^m)})$\; 
Compute $d_{min} = \min(d_{iter}, d_{exit}, d_{si})$ \;
\uIf{$d_{min} = d_{iter}$}{ 
  $x_p(t+\Delta t ) = x + \omega \times d_{iter}$\;
  $\omega_p(t+\Delta t ) = \omega $\;
  $t_p = 0$\;
} \uElseIf{$d_{min} = d_{exit}$}{
  $x_p(t+\Delta t ) = x_{min} \indi{\{  \omega = -1 \}} + x_{max} \indi{\{\omega =+1\}}$\;
  $ \omega_p(t+\Delta t ) = \omega $\;
  $t_p \leftarrow t_p - d_{exit}/c $\;
  Change the cell attributes or erase the particle if its arrives at the boundary of the domain\;
}\uElseIf{$d_{min} = d_{si}$}{
  $x \leftarrow x + \omega d_{si}$\;
  Draw $u_{[-1,1]} \sim \mathcal{U}_{\{[-1,1] \}} $\;
  $\omega \leftarrow \indi{\{u_{[-1,1] > 0}\}} - \indi{\{u_{[-1,1] < 0}\}}$ \;
  $t_p \leftarrow t_p - d_{si} / c $\;
}
$MCestimate(m) \leftarrow MCestimate(m) + \alpha \times (1- e^{-\sigma_a^m f^m d_{min}})$\;
$\alpha \leftarrow \alpha \times e^{-\sigma_a^m f^m d_{min}}$
}
\caption{ {\bf IMC classical} particle evolution in a rod geometry} 
\label{algo:IMC}
}
\end{algorithm}

\begin{algorithm}
{\modif
\KwData{quantization function {\it computeWithQuantization} defined in section~\ref{sec:method_algo} is loaded\;  
$x_p(t)$ is the position of the particle, $\omega_p(t)$ the angle of the particle, $t_p$ the remaining life time for the particle and the $\alpha_p(t)$ its weight. 
The particle is on cell $m$; this cell is represented by a segment $[x_{min}, x_{max}]$ ; the attributes $\sigma_a^m$ and $f^m$ of the cell are known from the table of mesh sizes $MCestimate$.}
Assign $\omega = \omega_p(t)$, $x = x_p(t)$, $\alpha = \alpha_p(t)$\;
\While{$t_p > 0$}{
$(x_f, \omega_f, t_s, state) \leftarrow computeWithQuantization(x, \omega, \sigma^m_a(1-f^m), c, x_{min}, x_{max}, t_p)$ \;
\uIf{$state = 0$}{ 
  $x_p(t+\Delta t ) = x + \omega t_pc$\;
  $\omega_p(t+\Delta t ) = \omega $\;
  $t_p = 0$\;
} \uElseIf{$state \ne 0$}{
  $x \leftarrow x_f $\;
  $ \omega \leftarrow \omega_f $\;
  $t_p \leftarrow t_p - t_s $\;
  Change the cell attributes or erase the particle if it arrives at the boundary of the domain\;
}
$MCestimate(m) \leftarrow MCestimate(m) + \alpha \times (1- e^{-\sigma_a^m f^m t_s  c})$\;
$\alpha \leftarrow \alpha \times e^{-\sigma_a^m f^m  t_s c}$
}
\caption{ {\bf \methodname{}} particle evolution in a rod geometry.} \label{algo:QIMC}
}
\end{algorithm}

\begin{algorithm}
{\modif 
Initialization:
for all cells $m$ initialize the attributes of the particles according to  the initial condition~: 
     $x_p(t_n) \sim \mathcal{U}_{ \{ [x_{min}^m, x_{max}^m] \}  }$ ;
     $\omega_p(t_n) \leftarrow \indi{\{u_{[-1,1] > 0}\}} - \indi{\{u_{[-1,1] < 0}\}}$ with $\mathcal{U}_{ \{ [-1, 1 \}  }$ ;
     $t_p = \Delta t$  ;
     $\alpha_p(t) = a T^4_{matter}(0, \cdots) (x_{max}^m, x_{min}^m) /N_{initial} $ \;
\While{$t_n < T_{final}$}{
Compute the Fleck factor \eqref{eq:fleck_beta} and the emission term $f_n a T_{matter}(t_n)$\;
For all remaining particles, take $t_p = \Delta t$ \;
For all cells $m$ initialize the attributes of particles for the emission term  \
     $x_p(t_n) \sim \mathcal{U}_{ \{ [x_{min}^m, x_{max}^m] \}  }$ ;
     $\omega_p(t_n) \leftarrow \indi{\{u_{[-1,1] > 0}\}} - \indi{\{u_{[-1,1] < 0}\}}$ with $\mathcal{U}_{ \{ [-1, 1] \}  }$ ;
     $t_p \sim \mathcal{U}_{ \{ [0, \Delta t] \} }$  ;
     $\alpha_p(t) = f^n_m \sigma_a^n \Delta t a T^4_{matter}(t_n) c (x_{max}^m - x_{min}^m) /N_{source} $ \;
Initialize the attributes of the boundary condition:
    $x_p(t_n) = 0$ ; 
    $\omega_p(t_n) = 1$ ;
    $t_p \sim \mathcal{U}_{ \{ 0, \Delta t \} }$ ;
    $\alpha_p(t) = 0.25 a c T^4_{matter}(\cdot, \textnormal{left border}) \Delta t /N_{boundary} $\;
Set the table $MCestimate$ to zero for all cell \;
For all particles use algorithm \ref{algo:IMC} or \ref{algo:QIMC} to obtain $MCestimate$ \;
Compute the new matter temperature using equation \eqref{eq:matiere_linearise}.
}
\caption{Resolution of the system \eqref{eq:IMC_rod} with a Monte Carlo method.} \label{algo:all_mc}
}
\end{algorithm}

 We use a model with two temperatures (radiative and matter)~: except  mention of the contrary, the term \textit{temperature} (noted $ T_{matter} $) will indicate the 
matter temperature.
This is a 1D benchmark in rod geometry (like the $S_N$ method~\cite{carlson_solution_1958} with $N=2$) with symmetry conditions on the top and bottom edges of the mesh.
 The values and units used are specified in the  table \ref{tab:valeursMarshak}.
We then solve the  system of equations \eqref{eq:IMC_rod} for $t \in [t^n, t^{n+1}[$ where $I^+(t, x) = u(t, x, \omega=+1)$ and $I^-(t, x) = u(t, x, \omega=-1)$ (see equation~\eqref{eq:transport}) {\modif the coupled matter interaction equation \eqref{eq:matiere}} and $f^n$ the Fleck factor {\modif definied by \eqref{eq:fleck_beta}}~:
\begin{equation}
\begin{aligned}
\frac{1}{c} \partial_t I^+ + \partial_x I^+ + \sigma^n f^n I^+ &= \sigma^n f^n \frac{a c T_{matter}^4(t^n)}{2} + \sigma^n (1-f^n) \frac{1}{2} (I^+ + I^-)   \\
\frac{1}{c} \partial_t I^- - \partial_x I^- + \sigma^n f^n I^- &= \sigma^n f^n \frac{a c T_{matter}^4(t^n)}{2} + \sigma^n (1-f^n) \frac{1}{2} (I^+ + I^-)   \\
C_V \partial_t T_{matter} &= \sigma^n f^n ( a c T_{matter}^4(t^n) - 2 \pi (I^+ + I^-))
\end{aligned}
\label{eq:IMC_rod}
\end{equation}

\begin{table*}[!htb]
	\centering
	\begin{tabular}{|c|c|c|c|c|}
		\hline
		$I$ & $erg . cm^{-2} . s^{-1}$ & $a$ & $ 7.56 \times 10^{-15} \ erg . cm^{-3} . K^{-4} $ \\ \hline
		$\Delta t$ & $4 \times 10^{-11} \ s$ & $d$ & $1.56 \times 10^{23} \  K^3 . g^{-1} . cm^2$ \\ \hline
		$\rho$ & $3 \  g . cm^{-3}$ & $c$ & $3 \times 10^{10} \ cm . s^{-1}$ \\ \hline
		$T_{matter}$ & $K$ & $T_{matter}(0,\cdot)$ & $11604 \ K$ \\ \hline
		$C_V$ & $8.6177 \times 10^{7} \ erg.g^{-1}.K^{-1}$ & $T_{matter}(\cdot, \text{left border})$ & $11604000 \ K$	 \\ \hline	 
	\end{tabular}
	\caption{\small \it Values and units used in the numerical simulation of the propagation of a Marshak-type wave in an opaque medium. \label{tab:valeursMarshak}}
\end{table*}

We analyze the wave profile at $1ns$, $5ns$ and $10ns$ using a time step of $\Delta t = 4 \times 10^{-11} s$.
To do this, we perform a run for the classical Monte Carlo method 
and a run with our method with
$50$ cells and $\nobj = 200$ (target number of particles by cell); we employ the local regularization method in~\cite{laguzet2020, LAGUZET2022111373} 
and compare the wave intensity to check for physical consistency.

The \methodname{} is tested 
with a temperature dependent opacity given by the formula~:~$\sigma = \rho \times d \times T_{matter}^{-3} cm^{-1}$ \cite{LAGUZET2022111373}.
The value used is computed at each iteration by the Fleck linearization method.
Note that the Fleck factor induces a scattering term also depending on the matter  temperature. This case illustrates the behavior of the method in a circumstance where the scattering values belong to different regimes. The results are presented in figure~\ref{fig:second_result_M}. 
The comparison with reference results shows good physical agreement, independent of the collision regime~: moreover the number of events per particle is substantially reduced (by a factor $1000$, cf. right axis in the right plot of  figure~\ref{fig:second_result_M}), together with the computation time.
Moreover, we notice that the computation time is no longer strictly proportional to the number of events as for the IMC classic method, which indicates that with this new method, the particle displacement phase is no longer the 
limiting phase in the computation time, but the treatment carried out between each tracking phase (emission and regulation of the particles for example) becomes important (the time increases with the number of particles remaining at the end of the iteration).
 
\FloatBarrier 

\section{Conclusion} \label{sec:conclusion}

We introduce the \methodname{} method to solve a computationally intensive multi-regime thermal radiative transport equation within an unifying framework. The method  is independent on any random walk assumptions to treat the high collision regime and 
relies on a offline computation followed by online sampling from a database. We check empirically that the smoothness assumptions underlying the method are, for the applications considered, of satisfactory quality; we next test the approach on a 1D benchmark and obtain physically coherent results while improving the computational time. 
This opens the perspective of future work on more complicated geometries and higher dimensional settings and beyond the 'gray' and isotropic approximation 
\footnote{\small Although this is not the focus of the present paper which works under the isotropic scattering approximation, using frequency groups could increase the memory requirements if the scattering is not isotropic any more. However 
	if the scattering remains isotropic, it seems to us that the frequency of each particle will 
	only relate to the $\sigma$ parameter whose range is fully considered in the dataset; so in this case the memory requirement will not necessarily increase. On the other hand the presence of multiple frequencies may induce some loss of precision which could necessitate to enlarge the dataset; the exact balance has to be fully asserted in upcoming works.} for the transport equations.
\begin{acknowledgments}
L.L. and G.T. acknowledge the support from their institutions.
\end{acknowledgments}


\begin{thebibliography}{10}
	
	\bibitem{pomraning1983equations}
	G~C Pomraning.
	\newblock {\em Equations of Radiation Hydrodynamics}.
	\newblock Pergamon Press, New York, 1983.
	
	\bibitem{review_Vlasov_Fokker_Planck_inertial_confinement12}
	A.G.R. Thomas, M.~Tzoufras, A.P.L. Robinson, R.J. Kingham, C.P. Ridgers, M.~Sherlock, and A.R. Bell.
	\newblock A review of {Vlasov-Fokker-Planck} numerical modeling of inertial confinement fusion plasma.
	\newblock {\em Journal of Computational Physics}, 231(3):1051--1079, 2012.
	\newblock Special Issue: Computational Plasma Physics.
	
	\bibitem{ATZENI2005153}
	S.~Atzeni, A.~Schiavi, F.~Califano, F.~Cattani, F.~Cornolti, D.~{Del Sarto}, T.V. Liseykina, A.~Macchi, and F.~Pegoraro.
	\newblock Fluid and kinetic simulation of inertial confinement fusion plasmas.
	\newblock {\em Computer Physics Communications}, 169(1):153--159, 2005.
	\newblock Proceedings of the Europhysics Conference on Computational Physics 2004.
	
	\bibitem{atmospheric_radiative_transfer_code_arts_v2_2011}
	P.~Eriksson, S.A. Buehler, C.P. Davis, C.~Emde, and O.~Lemke.
	\newblock Arts, the atmospheric radiative transfer simulator, version 2.
	\newblock {\em Journal of Quantitative Spectroscopy and Radiative Transfer}, 112(10):1551--1558, 2011.
	
	\bibitem{noebauer_monte_2019}
	Ulrich~M. Noebauer and Stuart~A. Sim.
	\newblock Monte {Carlo} radiative transfer.
	\newblock {\em Living Reviews in Computational Astrophysics}, 5(1):1, June 2019.
	
	\bibitem{castor2004radiation}
	John~I Castor.
	\newblock {\em Radiation Hydrodynamics}.
	\newblock Cambridge University Press, 2004.
	
	\bibitem{mihalas1999foundations}
	Dimitri Mihalas and Barbara Weibel-Mihalas.
	\newblock {\em Foundations of radiation hydrodynamics}.
	\newblock Courier Corporation, 1999.
	
	\bibitem{zeldovich1966physics}
	Y.B. Zel'dovich and Y.P. Raizer.
	\newblock {\em Physics of Shock Waves and High-temperature Hydrodynamic Phenomena}.
	\newblock Academic Press, 1966-1967.
	
	\bibitem{lapeyre1998methodes}
	B~Lapeyre, É~Pardoux, and R~Sentis.
	\newblock {\em Introduction to {Monte}-{Carlo} {Methods} for {Transport} and {Diffusion} {Equations}}.
	\newblock Oxford University Press, July 2003.
	
	\bibitem{densmore2005discrete}
	J.~D. Densmore, T.~J. Urbatsch, T.~M. Evans, and M.~W. Buksas.
	\newblock {Discrete Diffusion Monte Carlo for grey Implicit Monte Carlo simulations.}
	\newblock Technical report, Los Alamos National Laboratory, 2005.
	
	\bibitem{larsen1980diffusion}
	E.~W. Larsen.
	\newblock {Diffusion theory as an asymptotic limit of transport theory for nearly critical systems with small mean free paths}.
	\newblock {\em Annals of Nuclear Energy}, 7(4-5):249--255, 1980.
	
	\bibitem{FLECK1971313}
	J.A. Fleck and J.D. Cummings.
	\newblock {An implicit Monte Carlo scheme for calculating time and frequency dependent nonlinear radiation transport}.
	\newblock {\em Journal of Computational Physics}, 8(3):313 -- 342, 1971.
	
	\bibitem{brooks1989symbolic}
	E.~D. Brooks~III.
	\newblock {Symbolic Implicit Monte Carlo}.
	\newblock {\em Journal of Computational Physics}, 83(2):433--446, 1989.
	
	\bibitem{fleck1984random}
	J.A. Fleck and E.H. Canfield.
	\newblock {A random walk procedure for improving the computational efficiency of the implicit Monte Carlo method for nonlinear radiation transport}.
	\newblock {\em Journal of Computational Physics}, 54(3):508--523, 1984.
	
	\bibitem{giorla1987random}
	J.~Giorla and R.~Sentis.
	\newblock A random walk method for solving radiative transfer equations.
	\newblock {\em Journal of Computational Physics}, 70(1):145--165, 1987.
	
	\bibitem{gentile2001implicit}
	N.A. Gentile.
	\newblock {Implicit Monte Carlo diffusion - an acceleration method for Monte Carlo time-dependent radiative transfer simulations}.
	\newblock {\em Journal of Computational Physics}, 172(2):543--571, 2001.
	
	\bibitem{densmore2006hybrid}
	J.~D. Densmore, Todd~J. Urbatsch, T.~M. Evans, and M.~W. Buksas.
	\newblock {A hybrid transport-diffusion method for Monte Carlo radiative-transfert simulations}.
	\newblock {\em Journal of Computational Physics}, 222:485--503, 2007.
	
	\bibitem{pomraning1979}
	G.C. Pomraning and G.M. Foglesong.
	\newblock Transport-diffusion interfaces in radiative transfer.
	\newblock {\em Journal of Computational Physics}, 32(3):420--436, 1979.
	
	\bibitem{clouet2003hybrid}
	J-F. Clou{\"e}t and G.~Samba.
	\newblock {A Hybrid Symbolic Monte-Carlo method for radiative transfer equations}.
	\newblock {\em Journal of Computational Physics}, 188(1):139--156, 2003.
	
	\bibitem{COELHO201636}
	Pedro~J Coelho, Nicolas Crouseilles, Pedro Pereira, and Maxime Roger.
	\newblock Multi-scale methods for the solution of the radiative transfer equation.
	\newblock {\em Journal of Quantitative Spectroscopy and Radiative Transfer}, 172:36--49, 2016.
	\newblock Eurotherm Conference No. 105: Computational Thermal Radiation in Participating Media V.
	
	\bibitem{stamnes1981new}
	Knut Stamnes and Roy~A Swanson.
	\newblock A new look at the discrete ordinate method for radiative transfer calculations in anisotropically scattering atmospheres.
	\newblock {\em Journal of Atmospheric sciences}, 38(2):387--399, 1981.
	
	\bibitem{mobley2005interpretation}
	Curtis~D Mobley, Lydia~K Sundman, Curtiss~O Davis, Jeffrey~H Bowles, Trijntje~Valerie Downes, Robert~A Leathers, Marcos~J Montes, William~Paul Bissett, David~DR Kohler, Ruth~Pamela Reid, et~al.
	\newblock Interpretation of hyperspectral remote-sensing imagery by spectrum matching and look-up tables.
	\newblock {\em Applied Optics}, 44(17):3576--3592, 2005.
	
	\bibitem{lyapustin2011multiangle}
	Alexei Lyapustin, John Martonchik, Yujie Wang, Istvan Laszlo, and Sergey Korkin.
	\newblock Multiangle implementation of atmospheric correction (maiac): 1. radiative transfer basis and look-up tables.
	\newblock {\em Journal of Geophysical Research: Atmospheres}, 116(D3), 2011.
	
	\bibitem{martino2017automatic}
	Luca Martino, Jorge Vicent, and Gustau Camps-Valls.
	\newblock Automatic emulator and optimized look-up table generation for radiative transfer models.
	\newblock In {\em 2017 IEEE International Geoscience and Remote Sensing Symposium (IGARSS)}, pages 1457--1460. IEEE, 2017.
	
	\bibitem{modest2021radiative}
	Michael~F Modest and Sandip Mazumder.
	\newblock {\em Radiative heat transfer}.
	\newblock Academic press, 2021.
	
	\bibitem{book_quantization_measures}
	S.~Graf and H.~Luschgy.
	\newblock {\em Foundations of quantization for probability distributions}.
	\newblock Springer, 2007.
	
	\bibitem{laguzet_turinici_exit_law24}
	Laetitia Laguzet and Gabriel Turinici.
	\newblock Model free collision aggregation for the computation of escape distributions, 2024.
	\newblock arXiv:2403.10432, doi: https://doi.org/10.48550/arXiv.2403.10432.
	
	\bibitem{measure_quantization_turinici22}
	G.~Turinici.
	\newblock Huber-energy measure quantization.
	\newblock {\em submitted}, 2022.
	\newblock arXiv:2212.08162, doi 10.48550/arXiv.2212.08162.
	
	\bibitem{turinici_deep_2023}
	G.~Turinici.
	\newblock Deep {Conditional} {Measure} {Quantization}, 2023.
	\newblock arXiv:2301.06907, doi 10.48550/arXiv.2301.06907.
	
	\bibitem{https://doi.org/10.1002/eng2.12109}
	R.~Raghavan.
	\newblock Hitting time distributions for efficient simulations of drift-diffusion processes.
	\newblock {\em Engineering Reports}, 2(2):e12109, 2020.
	
	\bibitem{giorlaCEAn84}
	J.~Giorla and R.~Sentis.
	\newblock {Photonique Monte-Carlo dans les milieux opaques: m{\'e}thodes de Fleck avec "RANDOM WALK"}.
	\newblock {CEA-N} 2423, Novembre 1984.
	
	\bibitem{marshak1958}
	R.~E. Marshak.
	\newblock {Effect of Radiation on Shock Wave Behavior}.
	\newblock {\em The Physics of Fluids}, 1(1):24--29, 1958.
	
	\bibitem{MCCLARREN20089711}
	R.~G. McClarren and R.~B. Lowrie.
	\newblock The effects of slope limiting on asymptotic-preserving numerical methods for hyperbolic conservation laws.
	\newblock {\em Journal of Computational Physics}, 227(23):9711 -- 9726, 2008.
	
	\bibitem{laguzet2020}
	L.~Laguzet.
	\newblock Méthode locale pour l'échantionnage et la régulation des particules {Monte-Carlo} pour le transfert radiatif dans le code {FCI}2.
	\newblock {CEA-R} 6554, Janvier 2021.
	
	\bibitem{carlson_solution_1958}
	B.G. Carlson and G.I. Bell.
	\newblock Solution of the transport equation by the {Sn} method.
	\newblock Technical report, Los Alamos Scientific Lab., N. Mex., 1958.
	
	\bibitem{LAGUZET2022111373}
	L.~Laguzet and G.~Turinici.
	\newblock A cell-based population control of {Monte Carlo} particles for the global variance reduction for transport equations.
	\newblock {\em Journal of Computational Physics}, 467:111373, 2022.
	
\end{thebibliography}

\end{document}